\documentclass[a4paper,12pt]{article}
\usepackage[utf8]{inputenc}
\usepackage{amsmath}
\usepackage{verbatim}
\usepackage{tikz}
\usepackage{wrapfig}

\DeclareMathOperator{\sgn}{sgn}

\title{An asymptotic structure of the bifurcation boundary of the perturbed Painlev\'e-2 equation}
\author{O.~M.~Kiselev}

\begin{document}
\maketitle
\begin{abstract}
Solutions of the perturbed Painlev\'e-2 equation are typical for describing a dynamic bifurcation of soft loss of stability. The bifurcation boundary separates solutions of different types before bifurcation and before loss of stability. This border has a spiral structure. The equations of modulation of the bifurcation boundary depending on the perturbation are obtained. Both analytical and numerical results are given.
\end{abstract}

\section{Introduction}
\label{secIntro}

Here we construct an asymptotic perturbation theory for the Painlev\'e-2 equation in the form:
\begin{equation}
u''=-2u^3+xu-\varepsilon f(u,u',x),\quad 0<\varepsilon\ll1.
\label{eqPP2}
\end{equation}

Usually, perturbed Painlev\'e-2 equation appears in narrow layers when studying dynamic bifurcations of solutions of non-Autonomous differential equations. It was shown, for example, in the work \cite{Haberman1977}. A careful study of the relationship between the pitchfork  bifurcation and the Painlev\'e-2 equation was given in \cite{Maree1996}. The asymptotic behaviour of the Painlev\'e-2 equation due to the hard loss of stability loss was constructed in \cite{Kiselev2001} for the Painlev\'e-2 equation and in \cite{KiselevGlebov2003} for hard stability loss in the main resonance equation.

From the point of view of a general approach for dynamic bifurcations in second-order non-autonomous equations, similar questions are considered, for example, in \cite{Haberman2001}. A typical example is the doubling bifurcation in the theory of parametric autoresonance \cite{Kiselev-Glebov2007}. It should be noted here that in the listed works related to dynamic bifurcations \cite{Maree1996}, \cite{Haberman2001}, \cite{KiselevGlebov2003}, \cite{Kiselev-Glebov2007}, \cite{GlebovKiselevTarkhanov2017}, the Painlev\'e-2 equation plays an important role in the narrow transition layer and the constructed asymptotics of solutions are used for matching the parameters of solutions before and after the bifurcations. In this case, in a narrow layer, it is sufficient to link the parameters of the asymptotics before and after the transition with the parameters of the unperturbed Painlev\'e transcendent.

In general, the bifurcation structure of the hierarchy of Painlev\'e equations was discussed in \cite{KiselevSuleimanov1999}.  An approach to scaling limits in the Painlev\'e equations was considered, for example, in \cite{Kapaev1997Eng}.
\begin{figure}
\hspace{-1cm}
\includegraphics[scale=0.35]{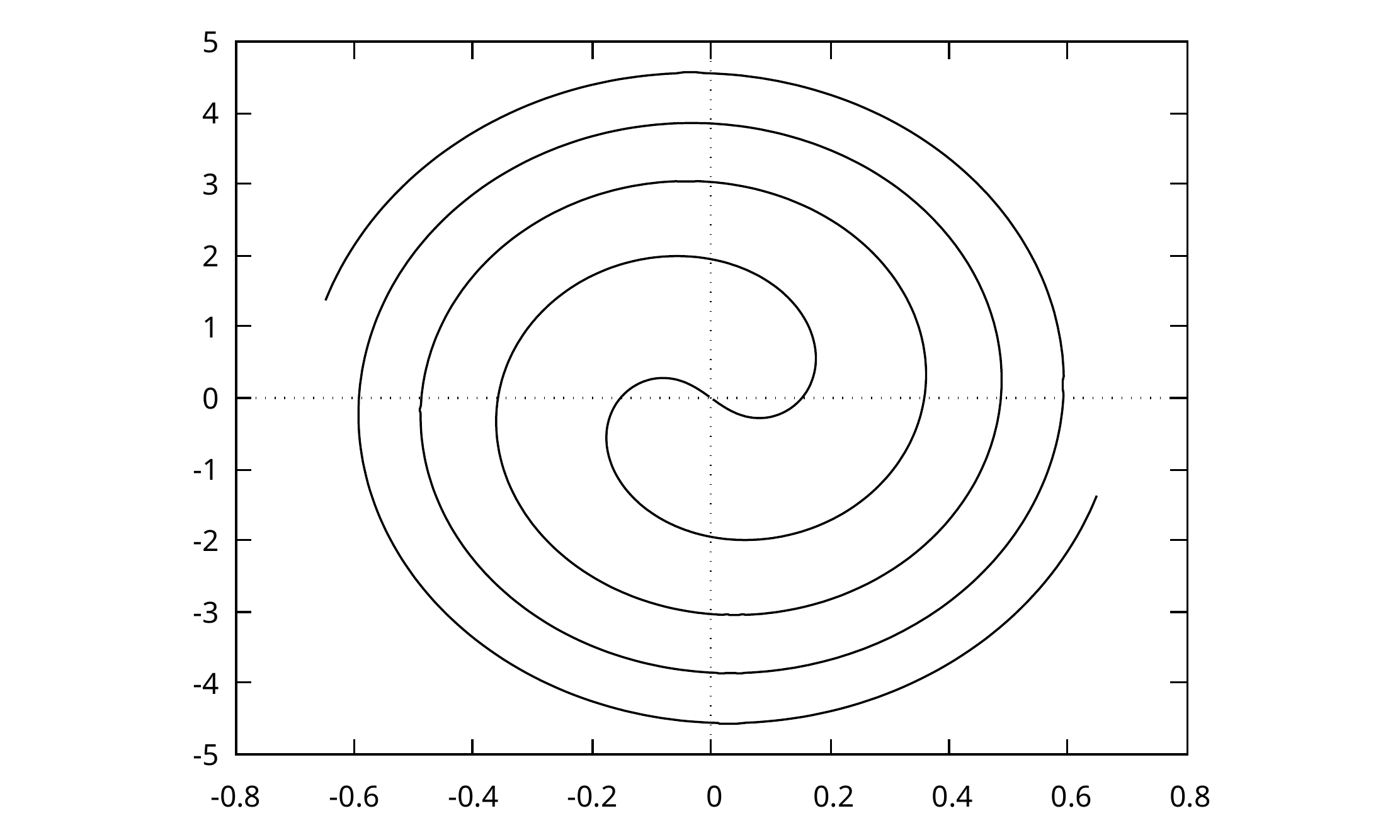}
\hspace{-1cm}
\includegraphics[scale=0.35]{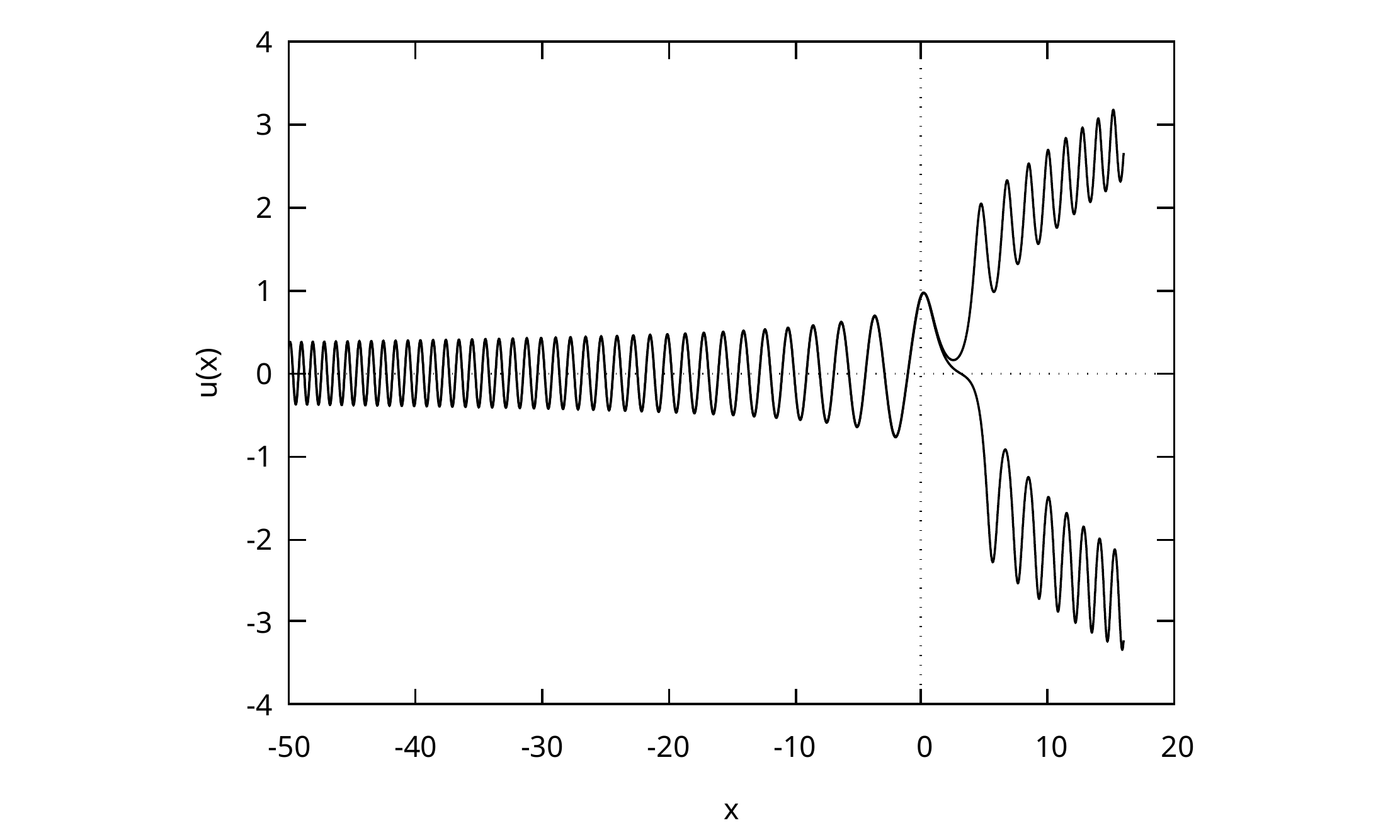}
\caption{In the left picture, the bifurcation boundary for the Painlev\'e-2 transcendent in the cross section of the phase space at $x=-50$. The right image shows two numerical solutions of the unperturbed Painlev\'e-2 equation with initial conditions taken near the bifurcation boundary. For $x<0$, the curves almost coincide; for $x>0$, they diverge as a result of soft loss of stability.}
\label{figDynamicalPichforkBifurcationOfPainleveTrancendent}
\end{figure}

A typical picture  for soft loss of stability of small oscillations for solutions of the Painlev\'e-2 equation has the form shown in figure \ref{figDynamicalPichforkBifurcationOfPainleveTrancendent}.
Here, on the left side of the figure, the solutions of the unperturbed Painlev\'e-2 equation are defined by asymptotics of the form:
\begin{equation}
u\sim \frac{\alpha}{\sqrt[4]{-x}}\sin\left(\frac{2}{3}(-x)^{3/2}+\frac{3}{4}\alpha^2\log(-x)+\phi\right),\quad x\to-\infty.
\label{formulaForLeftAsymptotics}
\end{equation}
The solution parameters are arbitrary constants $\alpha$ and $\phi$.

In the right part of the figure \ref{figDynamicalPichforkBifurcationOfPainleveTrancendent} solutions for $x\to\infty$ oscillate in the neighbourhood of branches of the function $ \pm\sqrt{x/2}$.

The boundary between the trajectories from the right side of the figure that the transition field through $x=0$ for the unperturbed Painlev\'e-2 equation is determined by the monodromy data. The relationship between monodromy data and the asymptotics of solutions of the Painlev\'e-2 equation is considered in the monograph \cite{ItsNovokshenov1986}. A detailed description is also available in \cite{ItsKapaevNovokshenovFokasEng}. The boundary line defining the solution for $x\to-\infty$, which for $x\to+\infty$ is determined by the asymptotic $ \pm\sqrt{x/2}$ connects the monodromy data and the asymptotics for $x\to - \infty$ (\cite{ItsKapaev1987Rus}):
\begin{equation}
\frac{3}{2}\alpha^2\log(2)-\frac{\pi}{4}-\arg\left(\Gamma\left(\frac{i\alpha^2}{2}\right)\right)=0.
\label{eqForBorderLine}
\end{equation}

When perturbation occurs, the bifurcation boundary is deformed, and the structure of the set of initial values that pass through various stable branches after bifurcation becomes more complex.

The effect of the disturbance is seen over a long period of time. Therefore, we can expect that it is sufficient to investigate the behaviour of the solution for large values of $|x|$. In this paper, we construct an asymptotic model for two parameters : the large parameter $|x|\gg1$ and the small parameter $0<\varepsilon\ll1$.

\section{Motivation and numerical results}
\label{secMotivation}

For a numerical study of the bifurcation boundary structure, it is convenient to study a family of trajectories released for a given value of the independent variable $-x_0\ll1$. Among the studied family of trajectories, two sets are obtained. There are the trajectories passing into the neighbourhood of $\sqrt{x/2}$ and into the neighbourhood of $-\sqrt{x/2}$ for $x>0$. Because of the need to explore large families, it is convenient to use parallel computing on the GPU.
\begin{figure}
\includegraphics[scale=0.5]{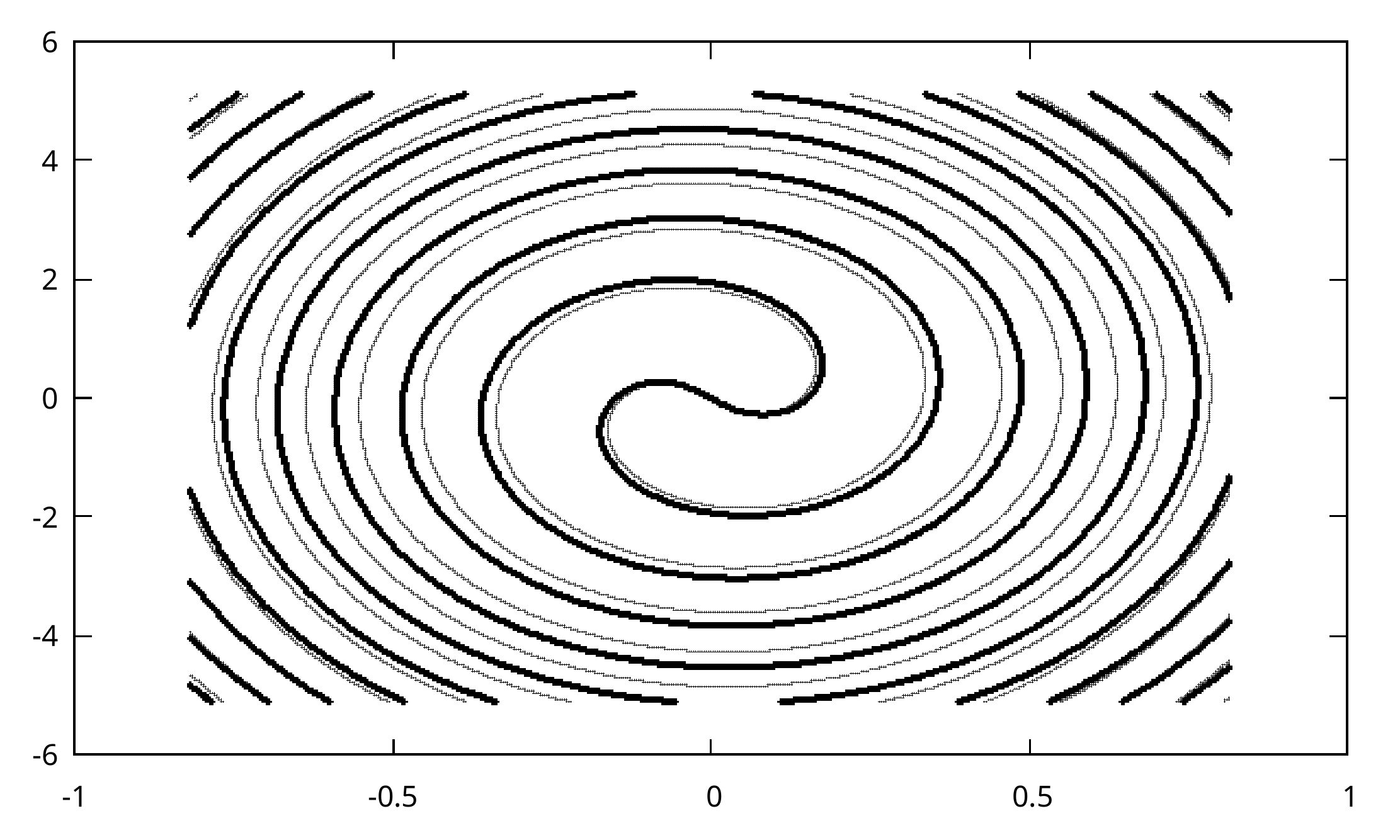}
\label{figBorderlineForPainleveAndPerturbedPainleve}
\caption{Here one can see the results of calculations of 2048x4096 trajectories by the Runge-Kutta method of the 4th order. The bifurcation boundary in the section of the phase space $(u,u',x)$ is given for $x=-50$. The bold curve corresponds to the Painlev\'e-2 equation, the thin curve corresponds to the perturbed equation (\ref{eqPP2}) with perturbation $f=u(u')^2$ for $\varepsilon=0.1$.} 
\end{figure}

The figure (\ref{figBorderlineForPainleveAndPerturbedPainleve}) shows two boundaries of bifurcation curves. There are the unperturbed Painleve-2 and the perturbed one. The curve for the perturbed equation is similar to the deformed curve for the unperturbed Painleve-2 equation.
\begin{figure}
\includegraphics[scale=0.3]{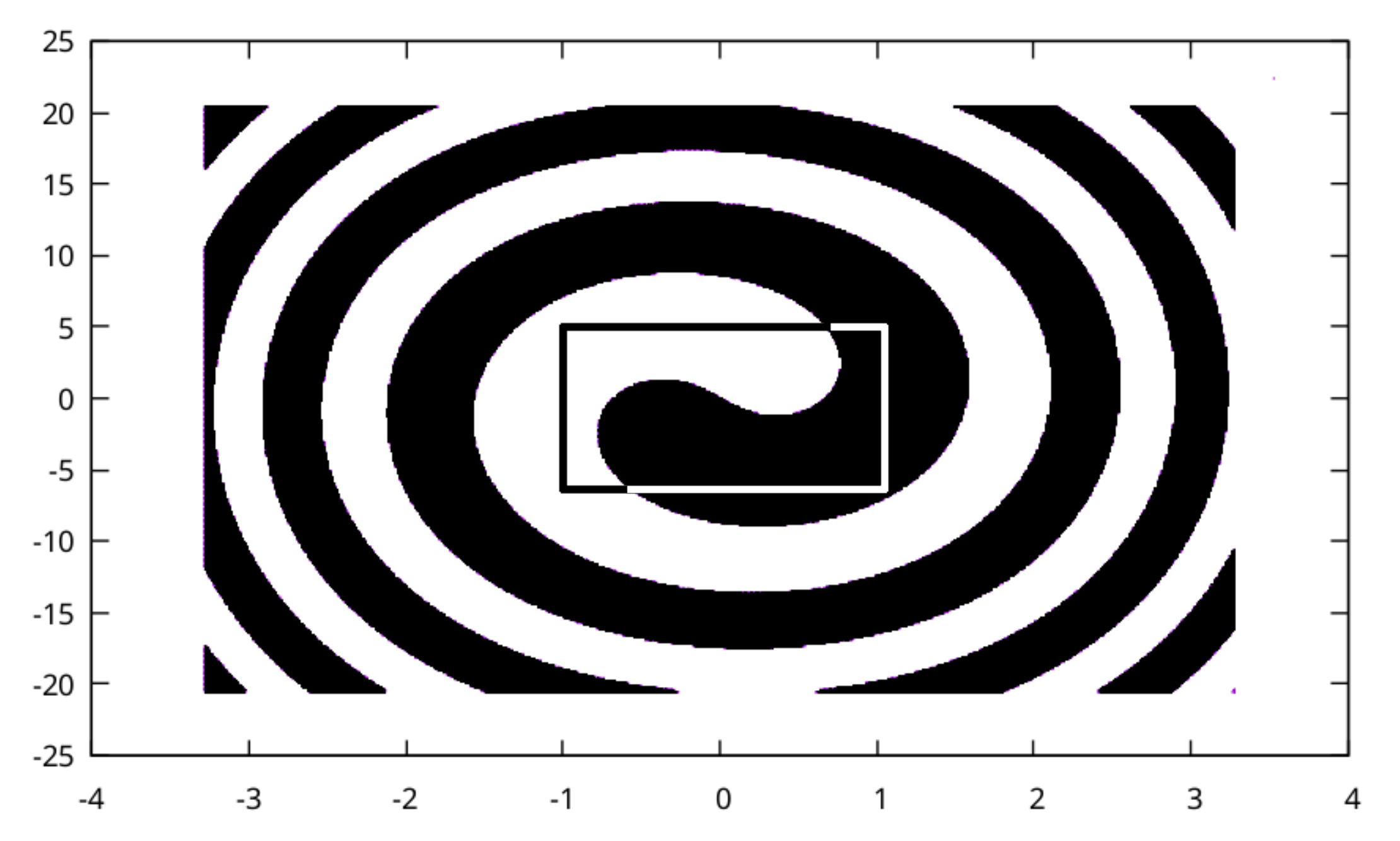}
\includegraphics[scale=0.3]{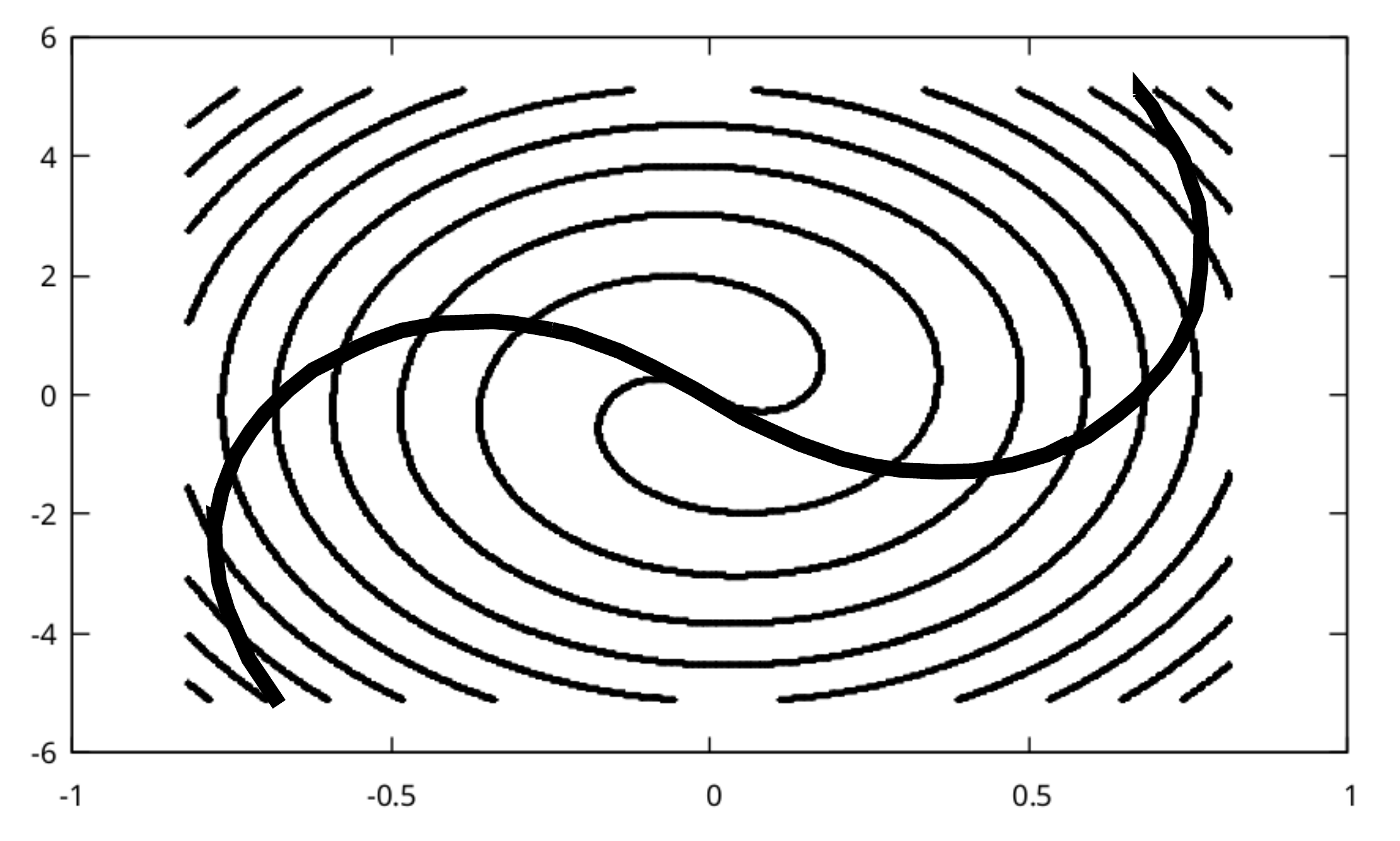}
\label{figBorderlineForPainleveAndDissipativePainleve}
\caption{Here the results of calculations of 2048x4096 trajectories by the Runge-Kutta method of the 4th order are shown. The set of trajectories in the cross section of the phase space $(u,u',x)$ at $x=-50$ for the equation (\ref{eqPP2}) with perturbation $f=u'$ at $\varepsilon=0.1$ is shown on the left. The dark part is the set of starting points of trajectories that, when passing through $x=0$, fall into the neighbourhood of $\sqrt{x/2}$. The bifurcation boundary of the set of trajectories is shown from the rectangle selected in the left image. In the right picture, the thin curves correspond to the Painlev\'e-2 equation, while the thick one corresponds to the perturbed equation (\ref{eqPP2}) with the perturbation $f=u'$ at $\varepsilon=0.1$.} 
\end{figure}
\par
Depending on the type of perturbation, the boundary deformation may be significant. For example, for a dissipative perturbation $f=u ' $ at $\varepsilon=0.1$ the results of calculations are shown in the figure (\ref{figBorderlineForPainleveAndDissipativePainleve}). The boundary of the perturbed equation is significantly distorted in comparison with the bifurcation boundary of the unperturbed Painleve-2 equation.

The goal of this work is to derive equations that determine the asymptotic behaviour of the bifurcation boundary for perturbations from a certain class.

\section{A formalism of perturbation theory}
\label{secFormalPerturbationTheory}
The asymptotics for the parameter $\varepsilon$ will be constructed as:
\begin{equation}
u(x,\varepsilon)\sim \sum_{k=0}^\infty \varepsilon^k u_k(x,\alpha,\phi).
\label{asymptoticsOfSolutionOfPerturbedPainleve}
\end{equation}
The main condition for representing corrections in the formula (\ref{asymptoticsOfSolutionOfPerturbedPainleve}) is uniform boundedness in $ \varepsilon$ for $x\to - \infty$.

As the primary term of the asymptotics of perturbation theory in $\varepsilon$, the  asymptotics of the Painlev\'e-2 transcendent for $x\to-\infty$, given in the formula (\ref{formulaForLeftAsymptotics}), is not suitable. In solving the perturbed equation, the parameters $\alpha$ and $\phi$ are functions of the independent variable $x$ and the parameter $\varepsilon$.

We assume that $\alpha$ and $\phi$ can be decomposed into a series by the parameter $\varepsilon$ and assume that the coefficients of this series depend on the slow variable $\xi=\varepsilon x$:
$$
\alpha\sim\sum_{k=0}^\infty \varepsilon^k\alpha_k(\xi),\quad
\phi\sim\sum_{k=0}^\infty \varepsilon^k\phi_k(\xi).
$$

Asymptotics of the primary term of the perturbed Painlev\'e transcendent:
$$
u_0(x,\alpha,\phi)\sim \frac{\alpha}{\sqrt[4]{-x}}\sin\left(\frac{2}{3}(-x)^{3/2}+\frac{3}{4}\int^{\xi/\varepsilon}\alpha^2(\zeta)\frac{d\zeta}{\zeta}+\phi\right),\quad x\to-\infty.
$$
Note that the integral in the argument of the function $\sin$ for the unperturbed Painlev\'e-2 equation gives the standard formula (\ref{formulaForLeftAsymptotics}).

To construct an asymptotic for the parameter $\varepsilon$, consider the equation for the first correction term for $\varepsilon$:
$$
u_1''=-6u_0^2u_1+xu_1-f(u_0,u_0',x) -2\alpha_0'\partial_{\alpha}u_0'-\phi_0' \partial_\phi u_0'.
$$

Here the function $u_1$ is the solution of the inhomogeneous linearized Painlev\'e-2 equation. The linearized Painlev\'e-2 equation has the form:
$$
v''=-6u^2 v+xv.
$$
Two linearly independent solutions to this equation can be obtained from the derivatives of the Painlev\'e-2 transcendent in the parameters $\alpha$ and $\phi$. Asymptotic behaviour of solutions of the linearized equation:
$$
v_1\sim\frac{1}{\sqrt[4]{-x}}\sin\left(\frac{2}{3}(-x)^{3/2}+\frac{3}{4}\alpha^2\log(-x)+\phi\right),\quad x\to-\infty,
$$
$$
v_2\sim \frac{1}{\sqrt[4]{-x}}\cos\left(\frac{2}{3}(-x)^{3/2}+\frac{3}{4}\alpha^2\log(-x)+\phi\right),\quad x\to-\infty.
$$
The Wronskian of these solutions can be calculated as follows:
$$
w=v_1v_2'-v_1'v_2\sim 1-\frac{3\alpha^2}{4\sqrt{-x^3}}\left(2+
\cos\left(\frac{4}{3}(-x)^{3/2}+\frac{3}{2}\alpha^2\log(-x)+\phi\right)\right).
$$
Since  Wronskian of linearly independent solutions is constant, then passing to the limit at $x\to - \infty$ in the right part of the asymptotic formula for  Wronskian we obtain:
$$
w=1.
$$
For a linearized Painlev\'e-2 equation with a modulated Painvel\'e transcendent, the $x$ asymptotic of two linearly independent solutions can be obtained by differentiating the asymptotics of the primary term in the parameters $\alpha$ and $\phi$. We denote these linearly independent solutions $u_\alpha$ and $u_\phi$, respectively:
$$
u_\alpha\sim \frac{1}{\sqrt[4]{-x}}\sin\left(\frac{2}{3}(-x)^{3/2}+\frac{3}{4}\int^{\xi/\varepsilon}\alpha^2(\zeta)\frac{d\zeta}{\zeta}+\phi\right),\quad x\to-\infty.
$$
$$
u_\phi\sim \frac{1}{\sqrt[4]{-x}}\cos\left(\frac{2}{3}(-x)^{3/2}+\frac{3}{4}\int^{\xi/\varepsilon}\alpha^2(\zeta)\frac{d\zeta}{\zeta}+\phi\right),\quad x\to-\infty.
$$

The solution of the equation for the first correction term can be represented using the formula:

\begin{eqnarray}
u_1=u_\alpha\int^x (f(u_0,u_0',y)-2\alpha_0'\partial_\alpha u_0'-2\phi_0'\partial_\phi u')u_\phi(y)dy-
\nonumber
\\
u_\phi\int^x (f(u_0,u_0',y)-2\alpha_0'\partial_\alpha u_0'-2\phi_0'\partial_\phi u')u_\alpha(y)dy.
\label{formForFirstCorection}
\end{eqnarray}
We calculate the integrals of the derivatives of the perturbed transcendent of Painlev\'e-2:
$$
\int^x \partial_\alpha u_0'u_\phi(y)dy\sim\int\cos^2\left(\frac{2}{3}(-y)^{3/2}+\frac{3}{4}\int^{\xi/\varepsilon}\alpha^2(\zeta)\frac{d\zeta}{\zeta}+\phi\right)dy\sim\frac{x}{2},
$$
$$
\int^x \partial_\phi u_0'u_\phi(y)dy=\frac{1}{2\alpha}(\partial_\phi u_0(x,\alpha,\phi))^2,
$$
$$
\int^x \partial_\alpha u_0'u_\alpha(y)dy=\frac{1}{2\alpha}(\partial_\alpha u_0(x,\alpha,\phi))^2;
$$
$$
\int^x \partial_\phi u_0'u_\alpha(y)dy=u_\alpha(x)\partial_\phi u_0(x,\alpha,\phi)-
\int^x u_\alpha(y)\partial_\phi u_0'(y,\alpha,\phi)dy\sim-\frac{x}{2}.
$$
Consequently, in the first correction term, secular terms may arise during integration.

A condition for discard the linear growth in the first correction term can be  obtained by averaging:
$$
\alpha_0'=\lim_{x\to-\infty}\frac{1}{x}\int^x f(u_0,u_0',y)u_\phi(y)dy,
$$
$$
\phi_0'=-\lim_{x\to-\infty}\frac{1}{x}\int^x f(u_0,u_0',y)u_\alpha(y)dy.
$$
The equations for higher corrections have the form:
$$
u_k''=6u_0^2u_k+xu_k-f_k(u_0,\dots,u_{k-1},u_0',\dots,u_{k-1}',x) -2\alpha_{k-1}'\partial_{\alpha}u_0'-2\phi_{k-1}' \partial_\phi u_0'.
$$
The averaging equations for the higher corrections are obtained similarly:
$$
\alpha_{k-1}'=\lim_{x\to-\infty}\frac{1}{x}\int^x (u_0,\dots,u_{k-1},u_0',\dots,u_{k-1}',x)u_\phi(y)dy,
$$
$$
\phi_{k-1}'=-\lim_{x\to-\infty}\frac{1}{x}\int^x (u_0,\dots,u_{k-1},u_0',\dots,u_{k-1}',x)u_\alpha(y)dy.
$$

\section{A deformation of the bifurcation boundary}
\label{secBorderDeformation}
\begin{figure}
\begin{center}
\includegraphics[scale=0.3]{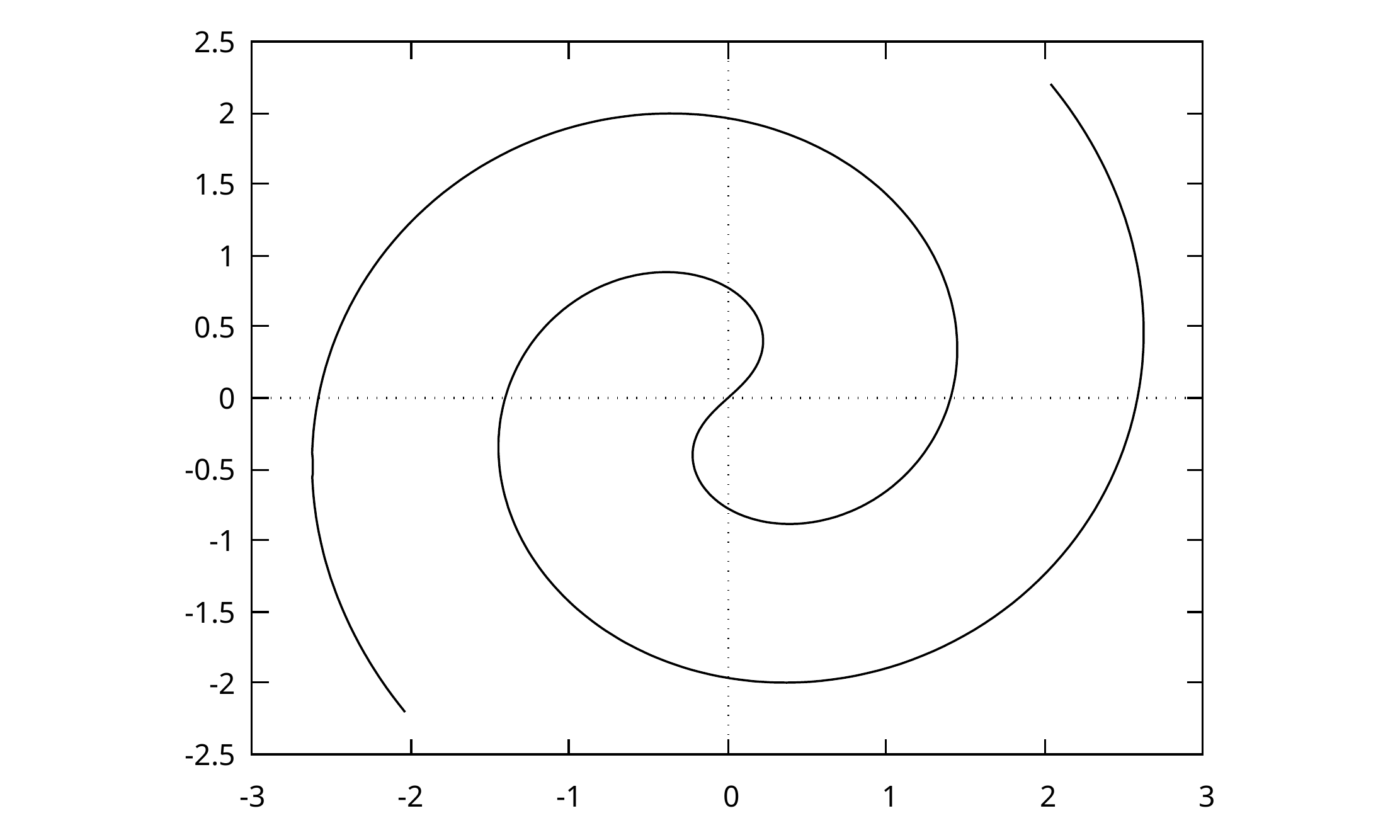}
\end{center}
\label{bifurcationBorderForPainleve2}
\caption{Bifurcation boundary for the parameters of the Painlev\'e-2 transcendent in the polar coordinate system, here $r$ is the distance from the coordinate axis, $((3/2) r^2\log(2)-\pi/4-\arg(\Gamma(ir^2/2)$ is the angle relative to the abscissa axis.}
\end{figure}
In the theory of Painlev\'e transcendent, it is known that for $x\to\infty$ solutions, two families can be divided according to the asymptotic behaviour. These families and their relation to monodromy data were established in the already mentioned works \cite{ItsKapaev1987Rus} and \cite{Belogrudov1997}. The sign of this expression
\begin{equation}
\kappa=\sin\left(((3/2)\alpha^2\log(2)-\pi/4-\arg(\Gamma(i\alpha^2/2)-\phi)\right).
\label{formulaForSignum}
\end{equation}
defines a bifurcation transition for $x\to\infty$:
\begin{eqnarray}
u\sim -\sgn(\kappa)\sqrt{\frac{x}{2}}.
\label{asymptoticsGrowingForPainleve}
\end{eqnarray}
The boundary in terms of $\alpha,\phi$ is shown in figure \ref{bifurcationBorderForPainleve2}.

The value of the parameters $\alpha (0)$ and $\phi (0)$ determines the type of bifurcation transition for the perturbed equation. Namely, the well-known $\alpha(0)$ and $\phi(0)$ can be used to determine the sign of the expression (\ref{formulaForSignum}).
\par
Let's consider the inverse problem. Determine the deformation of the bifurcation boundary for some $ \xi<0$. To do this, let's parametrize the initial curve in the polar coordinate system:
$$
\alpha(r,0)=r,\quad
\phi(r,0)=((3/2)r^2\log(2)-\pi/4-\arg(\Gamma(ir^2/2).
$$
Then, for $-x\gg1$, the bifurcation boundary in the $x$ section of the phase space for small $\varepsilon$ will be defined by the formulas:
\begin{eqnarray*}
u(x,\alpha(r,\xi),\phi(r,\xi))\sim
&
\frac{\alpha(r,\xi)}{\sqrt[4]{-x}}\sin\left(\frac{2}{3}(-x)^{3/2}+\frac{3}{4}\int^{\xi/\varepsilon}\alpha^2(r,\zeta)\frac{d\zeta}{\zeta}+\phi(r,\xi)\right),
\\
u'(x,\alpha(r,\xi),\phi(r,\xi))
\sim
&
\\
\alpha(r,\xi)
&
\sqrt[4]{-x}
\cos\left(\frac{2}{3}(-x)^{3/2}+\frac{3}{4}\int^{\xi/\varepsilon}\alpha^2(r,\zeta)\frac{d\zeta}{\zeta}+\phi(r,\xi)\right).
\end{eqnarray*}

\section{The Painlev\'e-2 equation with a dissipative additive}
\label{secDissipativePainleve2}

\begin{figure}
\includegraphics[scale=0.5]{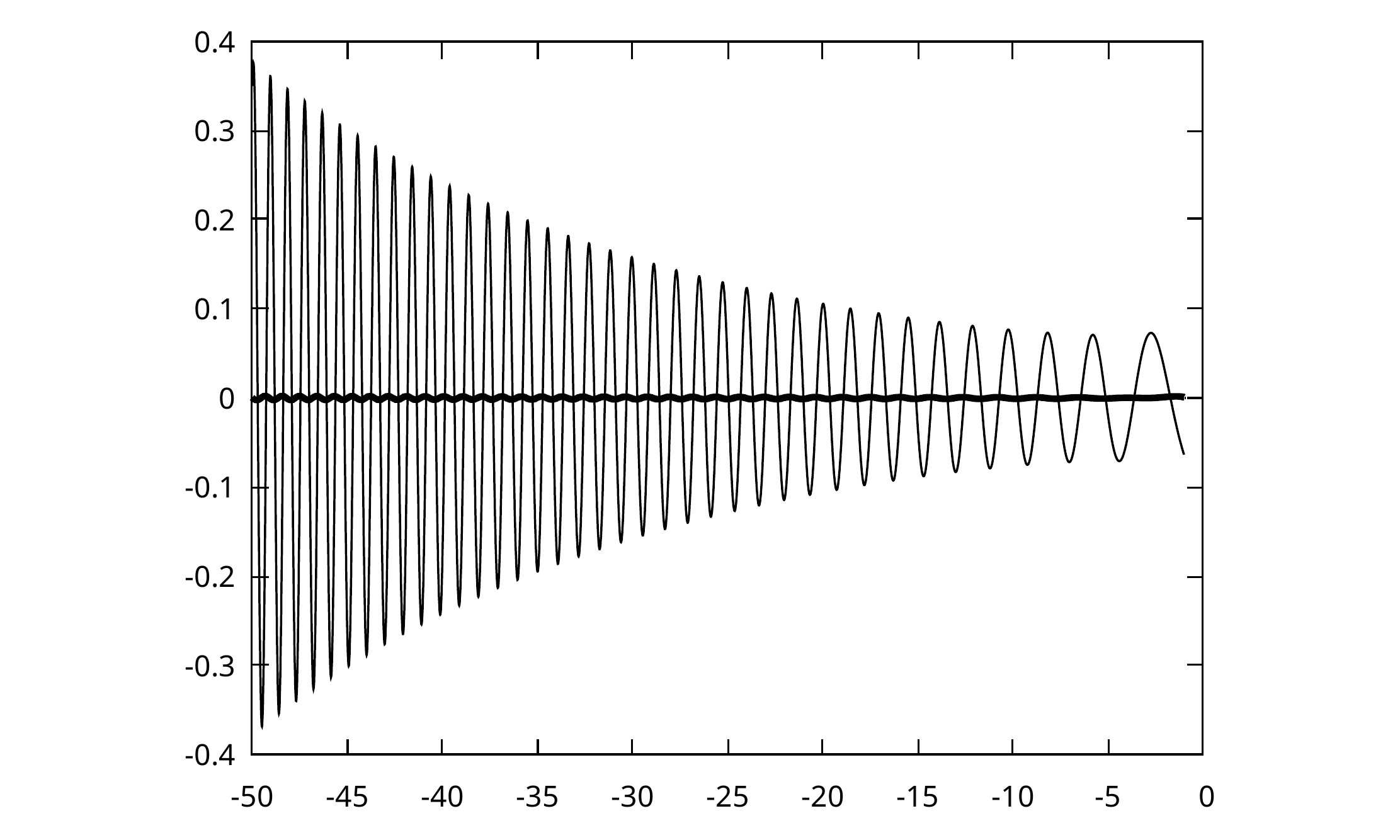}
\label{figNumDP2AndDiffBetweenAsymp}
\caption{ The numerical solution of the equation (\ref{equationPainleve2WithDissipation}) on the figure practically coincides with the constructed asymptotic solution. Therefore, the figure shows the numerical solution, which is  a thin line, and the difference between the numerical solution and the constructed asymptotic solution, which is  a bold line near the abscissa axis.}
\end{figure}

Here we consider an example of using the theory of non-integrable perturbations of the second transcendent of Painlev\'e developed above. Consider the Painlev\'e-2 equation with a small dissipative term:
\begin{equation}
u''=-2u^3+xu-\varepsilon u'.
\label{equationPainleve2WithDissipation}
\end{equation}

The linearization in the neighbour  of zero for this equation leads to the Airy equation with a dissipative term:
\begin{eqnarray}
y''=xy-\varepsilon y'.
\label{eqAiryWithDissipation}
\end{eqnarray}
The solution of the equation (\ref{eqAiryWithDissipation}) can be represented as an integral:
$$
y=\frac{1}{\pi}\int_0^\infty\cos\left(kx+\frac{k^3}{3}\right)e^{-\varepsilon k^2/2}dk.
$$
Asymptotic solution of this function for $x\to - \ infty$:
$$
y\sim\frac{e^{-\varepsilon x/2}}{\sqrt{2\pi}}\frac{\cos\left(\frac{2}{3}(-x)^{3/2}\right)}{(-x)^{1/4}}.
$$
Therefore, the amplitude of the solution oscillations decreases when the independent variable on the negative half-axis changes towards larger values, i.e. towards the point $x=0$. Similar results can be expected for solutions of the small-amplitude Painlev\'e-2 equation.

For solutions of the unperturbed Painlev\'e-2 equation, formulas are known about the relationship between the parameters of the asymptotic solutions of the Painlev\'e-2 equation and the monodromy data for it, see  \cite{ItsKapaev1987Rus}, \cite{Belogrudov1997}. Here, for simplicity of calculations, solutions from \cite{ItsKapaev1987Rus}, \cite{Belogrudov1997} are considered only in those regions of the independent variable $x$ where they are bounded or small.
\begin{figure}
\begin{center}\includegraphics[scale=1]{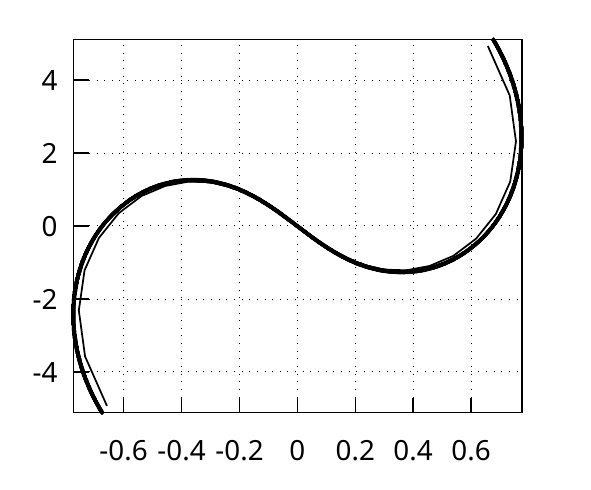}
\end{center}
\label{figDP2NumAndAsympBorder}
\caption{Here one can see the cross-section of the $u, u'$ bifurcation boundary at $x=-50$ for solutions of the perturbed Painlev\'e-2 equation with small dissipation (\ref{equationPainleve2WithDissipation}) at $\varepsilon=0.1$. the bold line is the boundary obtained numerically from 2048x4096 trajectories, with the beginning at $x=-50$. A thin line is a boundary calculated from perturbation theory.}
\end{figure}
According to the calculations in the Section \ref {secFormalPerturbationTheory}  the equation for modulating the parameters of the Painlev\'e trencendent asymptotic is:
$$
\alpha_0'=-\lim_{x\to-\infty}\frac{1}{x}\int^x u'(y,\alpha,\phi)u_\phi(y)dy
$$
Substituting the asymptotics to the right-hand side of this formula gives:
$$
\alpha_0'=-\alpha\lim_{x\to-\infty}\frac{1}{x}\int^x \cos^2\left(\frac{2}{3}(-y)^{3/2}+\frac{3}{4}\int^{\xi/\varepsilon}\alpha^2(\zeta)\frac{d\zeta}{\zeta}+\phi\right)dy.
$$
Integrating we get:
$$
\alpha_0'\sim -\frac{1}{2}\alpha_0.
$$
The equation for the modulation of $\phi_0$ is:
$$
\phi_0=\lim_{x\to-\infty}\frac{1}{x}\int^x u'(y,\alpha,\phi)u_\alpha(y)dy
$$
after substituting the main terms, the asymptotic gives:
\begin{eqnarray*}
\phi_0'\sim\lim_{x\to-\infty}\frac{1}{2x}\int^x\sin\left(\frac{4}{3}(-y)^{3/2}+\frac{3}{2}\int^{\xi/\varepsilon}\alpha^2(\zeta)\frac{d\zeta}{\zeta}+2\phi\right)
dy.
\end{eqnarray*}
Going to the limit leads to the equation
$$
\phi_0'=0.
$$
Then the asymptotic behaviour of the primary term of the Painlev\'e-2 transcendent with small dissipation has the form:
$$
u_0\sim a\frac{ e^{-\varepsilon x/2}}{\sqrt[4]{-x}}\sin\left(\frac{2}{3}(-x)^{3/2}+\frac{3 a^2}{4}\int^{x} \frac{e^{-\varepsilon z}}{z}dz+p\right),\quad x\to-\infty.
$$
Here $a$ and $p$ are solution parameters.

A comparison of the constructed asymptotics and the numerical solution for $ \alpha=1, p=0$ is shown in figure \ref{figNumDP2AndDiffBetweenAsymp}.
\par
The constructed asymptotics allows us to obtain a bifurcation boundary for any value of $-x\gg1$. In particular, figure \ ref{figDP2NumAndAsympBorder} shows a comparison between the bifurcation boundary obtained by numerical analysis of 2048x4096 trajectories starting at $x=-50$ and the bifurcation boundary calculated using the asymptotic formula. Figure \ref{figDP2NumAndAsympBorder} shows that these curves are close.

\section{Perturbed Painlev\'e-2 equation with non-linear perturbation}
\label{secNonlinearPerturbedP2}
\begin{figure}[ht]
\begin{center}
\includegraphics[scale=0.3]{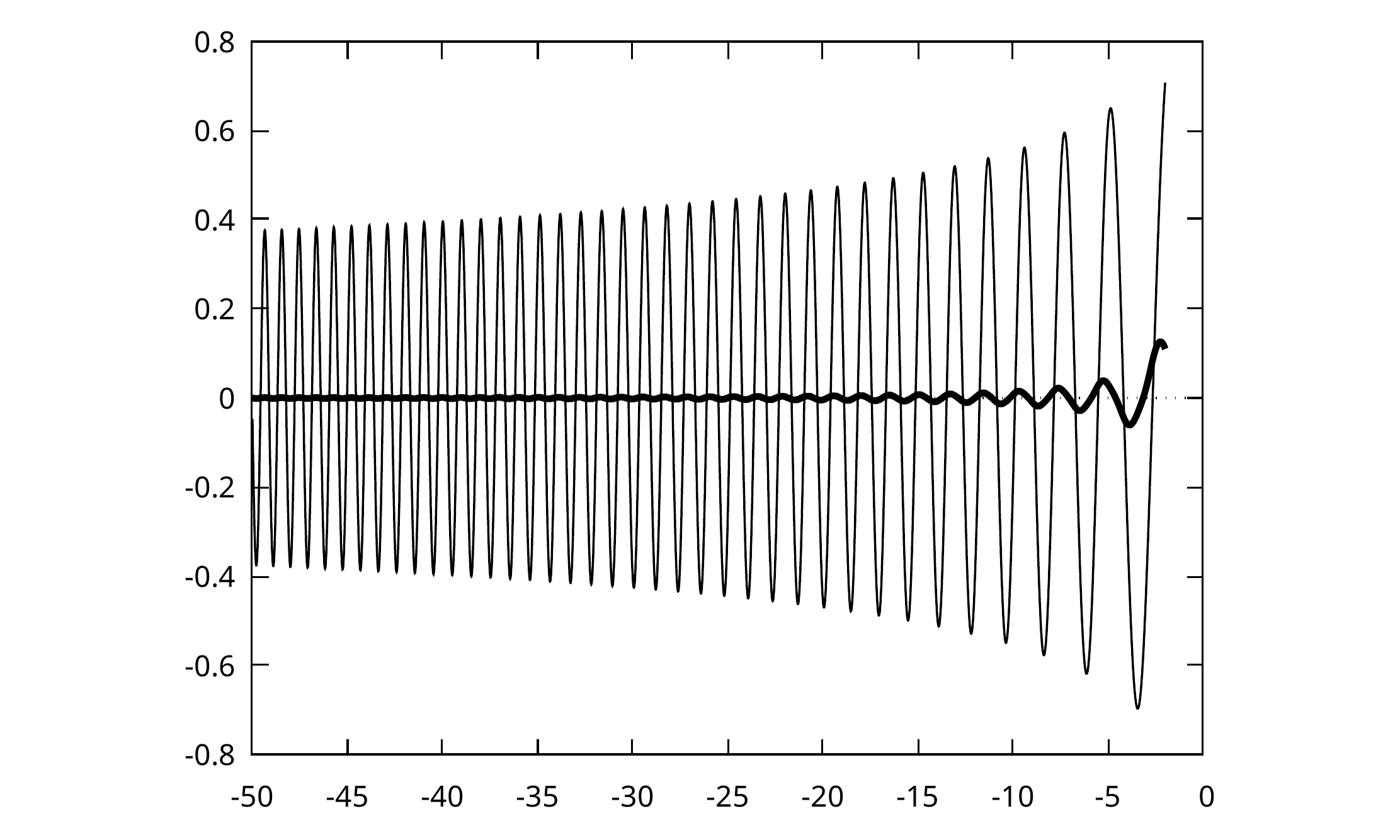}
\end{center}
\label{figNumNonlinPertP2AndDiffBetweenAsymp}
\caption{ The numerical solution of the equation (\ref{eqP2WithNonlinearPerturbation}) on the figure  practically coincides with the constructed asymptotic solution. Therefore, the figure shows the numerical solution, which is  a thin line, and the difference between the numerical solution and the constructed asymptotic solution, which is a bold line near the abscissa axis.}
\end{figure}

Here we consider another example of the perturbed Painlev\'e-2 equation:
\begin{equation}
u"=-2u^3+xu-\varepsilon (u')^2 u .
\label{eqP2WithNonlinearPerturbation}
\end{equation}
For this equation, modulating the parameter $\alpha$ is follow:
\begin{eqnarray*}
\alpha_0'\sim
&
\lim_{x\to-\infty}\frac{1}{x}\int^x
\alpha_0^3\cos^3\left(\frac{2}{3}(-y)^{3/2}+\frac{3}{4}\int^{\xi/\varepsilon}\alpha^2(\zeta)\frac{d\zeta}{\zeta}+\phi\right)
\times
\\
&
\times
\sin\left(\frac{2}{3}(-y)^{3/2}+\frac{3}{4}\int^{\xi/\varepsilon}\alpha^2(\zeta)\frac{d\zeta}{\zeta}+\phi\right)dy
\end{eqnarray*}
Going to the limit gives:
$$
\alpha_0'=0.
$$
Equation for modulation $\phi$ is:
\begin{eqnarray*}
\phi_0'\sim
&
\lim_{x\to-\infty}\frac{1}{x}\int^x
\alpha_0^3\cos^2\left(\frac{2}{3}(-y)^{3/2}+\frac{3}{4}\int^{\xi/\varepsilon}\alpha^2(\zeta)\frac{d\zeta}{\zeta}+\phi\right)
\times
\\
&
\times
\sin^2\left(\frac{2}{3}(-y)^{3/2}+\frac{3}{4}\int^{\xi/\varepsilon}\alpha^2(\zeta)\frac{d\zeta}{\zeta}+\phi\right)dy
\end{eqnarray*}
Going to the limit gives:
$$
\phi'\sim-\frac{1}{8}\alpha^3.
$$

\begin{figure}
\begin{center}
\includegraphics[scale=0.7]{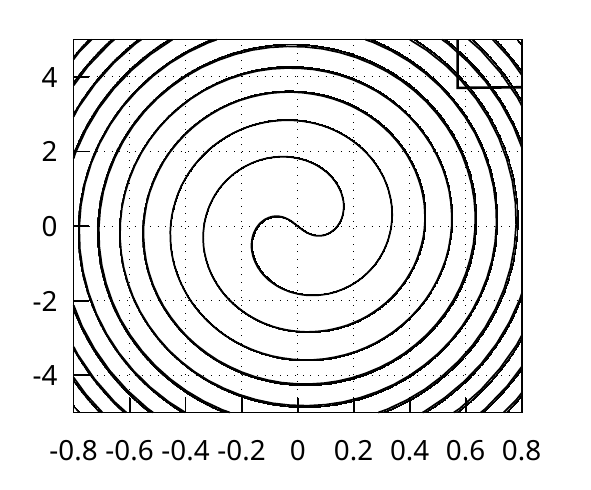}
\includegraphics[scale=0.7]{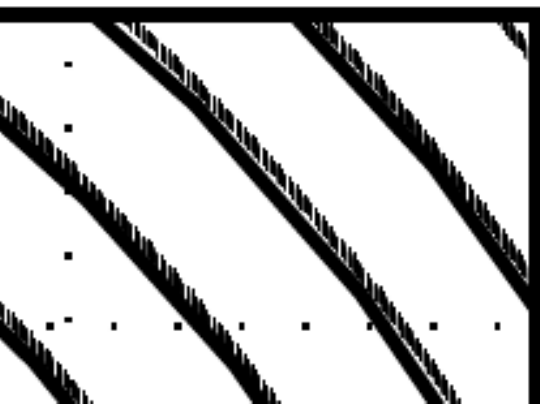}
\label{figNumAndAsympBorderLineNonlinPertP2}
\end{center}
\caption{Here one can see the cross-section of the $u,u'$ bifurcation boundary at $x=-50$ for solutions of the non-linearly perturbed Painlev\'e-2 equation (\ref{eqP2WithNonlinearPerturbation}) at $\varepsilon=0.1$. The boundary obtained numerically for$2048$x$ 4096 $ trajectories, with the beginning at $x=-50$ , and the boundary calculated from perturbation theory almost coincide. Differences can be observed away from the center. The rectangle highlighted in the left drawing is enlarged in the right drawing. In the right drawing, the boundary obtained numerically corresponds to short vertical dashes, the boundary obtained by perturbation theory is indicated by continuous lines.}
\end{figure}

That is, the perturbation leads to a shift:
\begin{equation}
u_0\sim \frac{\alpha}{\sqrt[4]{-x}}\sin\left(\frac{2}{3}(-x)^{3/2}+\frac{3 a^2}{4}\log(-x)-\frac{1}{8}\alpha^3\varepsilon x+p\right),\quad x\to-\infty.
\label{formulaForAsymptoticOfNonlinearPerturbedPainleve2}
\end{equation}
Here $\alpha$ and $p$ are solution parameters.

The figure 7 %\ref{figNumNonlinPertP2AndDiffBetweenAsymp} 
shows the numerical solution of the equation (\ref{eqP2WithNonlinearPerturbation}) and the difference between the numerical solution and the constructed asymptotics.

The figure 8 
%\ref{figNumAndAsympBorderLineNonlinPertP2} 
shows a cross-section for $x=-50$ of the bifurcation boundary for solutions of the perturbed Painlev\'e-2 equation (\ref{eqP2WithNonlinearPerturbation}). The figure shows that the boundary constructed from numerical results and the boundary constructed from asymptotic formulas are close.

\section{Conclusion}
The equations for the parameters of the asymptotic behaviour of the Painlev\'e-2 transcendent at $x \to-\infty$ derived in \ref{secFormalPerturbationTheory} allow us to obtain a formula for the bifurcation boundary of solutions for a perturbed equation with a soft loss of stability in the neighbourhood of $x=0$. This makes it possible to divide the solutions of the perturbed equation into solutions close to $\sqrt{x/2}$ and close to $-\sqrt{x/2}$ for $x\to \infty$. The results are illustrated by computing perturbations of various classes in the sections \ref{secDissipativePainleve2} and \ref{secNonlinearPerturbedP2}.

\end{document}